%Paper: hep-th/9409061
%From: cvj@guinness.ias.edu (Clifford Johnson)
%Date: Sun, 11 Sep 94 19:54:08 EDT
%Date (revised): Sun, 18 Sep 94 16:29:08 EDT

%\input cvjmac.tex

%%%%%%%%%%%%%%%%  preprint and letter macro  %%%%%%%%%%%%%%%%%

\hsize=6.0truein
\vsize=8.5truein
\voffset=0.25truein
\hoffset=0.1875truein%would be 0.25 except our laser printer is off by 1/16in
\tolerance=1000
\hyphenpenalty=500
\def\monthintext{\ifcase\month\or January\or February\or
   March\or April\or May\or June\or July\or August\or
   September\or October\or November\or December\fi}

%%%%%%%%%%%%%%%%%  Twelve point text font  %%%%%%%%%%%%%%%%%%%

\font\tenrm=cmr10 scaled \magstep1   \font\tenbf=cmbx10 scaled \magstep1
\font\sevenrm=cmr7 scaled \magstep1  
\font\fiverm=cmr5 scaled \magstep1   

\font\teni=cmmi10 scaled \magstep1   \font\tensy=cmsy10 scaled \magstep1
\font\seveni=cmmi7 scaled \magstep1  \font\sevensy=cmsy7 scaled \magstep1
\font\fivei=cmmi5 scaled \magstep1   \font\fivesy=cmsy5 scaled \magstep1

\font\tentt=cmtt10 scaled \magstep1
\font\tenit=cmti10 scaled \magstep1
\font\tensl=cmsl10 scaled \magstep1

\def\twelvepoint{\def\rm{\fam0\tenrm}
   \textfont0=\tenrm \scriptfont0=\sevenrm \scriptscriptfont0=\fiverm
   \textfont1=\teni  \scriptfont1=\seveni  \scriptscriptfont1=\fivei
   \textfont2=\tensy \scriptfont2=\sevensy \scriptscriptfont2=\fivesy
   \textfont\itfam=\tenit \def\it{\fam\itfam\tenit}
   \textfont\ttfam=\tentt \def\tt{\fam\ttfam\tentt}
   \textfont\bffam=\tenbf \def\bf{\fam\bffam\tenbf}
   \textfont\slfam=\tensl \def\sl{\fam\slfam\tensl} \rm
   %Essentially I changed all dimensions to 1.2 times as large as in plain tex
   \hfuzz=1pt\vfuzz=1pt%much more than plain tex's value
   \setbox\strutbox=\hbox{\vrule height 10.2pt depth 4.2pt width 0pt}
   \parindent=24pt\parskip=1.2pt plus 1.2pt
   \topskip=12pt\maxdepth=4.8pt\jot=3.6pt
   \normalbaselineskip=14.4pt\normallineskip=1.2pt
   \normallineskiplimit=0pt\normalbaselines
   \abovedisplayskip=13pt plus 3.6pt minus 5.8pt
   \belowdisplayskip=13pt plus 3.6pt minus 5.8pt
   \abovedisplayshortskip=-1.4pt plus 3.6pt
   \belowdisplayshortskip=13pt plus 3.6pt minus 3.6pt
   %plain tex's value for belowdisplayshortskip looked terrible
   \topskip=12pt \splittopskip=12pt
   \scriptspace=0.6pt\nulldelimiterspace=1.44pt\delimitershortfall=6pt
   \thinmuskip=3.6mu\medmuskip=3.6mu plus 1.2mu minus 1.2mu
   \thickmuskip=4mu plus 2mu minus 1mu%reduced these plain tex values
   \smallskipamount=3.6pt plus 1.2pt minus 1.2pt
   \medskipamount=7.2pt plus 2.4pt minus 2.4pt
   \bigskipamount=14.4pt plus 4.8pt minus 4.8pt}

\twelvepoint

%%%%%%%%%%%%%%%%% Definitions for Preprints %%%%%%%%%%%%%%%%%%

% title page title font

\font\titlerm=cmr10 scaled \magstep3
\font\titlerms=cmr10 scaled \magstep1 %\font\titlermss=cmr8
\font\titlei=cmmi10 scaled \magstep3  %math italic for title
\font\titleis=cmmi10 scaled \magstep1 %\font\titleiss=cmmi8
\font\titlesy=cmsy10 scaled \magstep3 	%math symbols for title
\font\titlesys=cmsy10 scaled \magstep1  %\font\titlesyss=cmsy8
\font\titleit=cmti10 scaled \magstep3	%text italic for title
\skewchar\titlei='177 \skewchar\titleis='177 %\skewchar\titleiss='177
\skewchar\titlesy='60 \skewchar\titlesys='60 %\skewchar\titlesyss='60

\def\titlefont{\def\rm{\fam0\titlerm}% switch to title font
   \textfont0=\titlerm \scriptfont0=\titlerms %\scriptscriptfont0=\titlermss
   \textfont1=\titlei  \scriptfont1=\titleis  %\scriptscriptfont1=\titleiss
   \textfont2=\titlesy \scriptfont2=\titlesys %\scriptscriptfont2=\titlesyss
   \textfont\itfam=\titleit \def\it{\fam\itfam\titleit} \rm}

% title page macros

\def\preprint#1{\baselineskip=19pt plus 0.2pt minus 0.2pt \pageno=0
   \begingroup%use with \draft or \date to end group
   \nopagenumbers\parindent=0pt\baselineskip=14.4pt\rightline{#1}}
\def\title#1{
   \vskip 0.9in plus 0.45in
   \centerline{\titlefont #1}}
\def\secondtitle#1{}%set up this some time
\def\author#1#2#3{\vskip 0.9in plus 0.45in
   \centerline{{\bf #1}\myfoot{#2}{#3}}\vskip 0.12in plus 0.02in}
\def\secondauthor#1#2#3{}%set up this some time
\def\addressline#1{\centerline{#1}}
\def\abstract{\vskip 0.7in plus 0.35in
	\centerline{\bf Abstract}
	\smallskip}
\def\finishtitlepage#1{\vskip 0.8in plus 0.4in
   \leftline{#1}\supereject\endgroup}

\def\date#1{\finishtitlepage{#1}}

\def\nolabels{\def\eqnlabel##1{}\def\eqlabel##1{}\def\figlabel##1{}%
	\def\reflabel##1{}}
\def\writelabels{\def\eqnlabel##1{%
	{\escapechar=` \hfill\rlap{\hskip.11in\string##1}}}%
	\def\eqlabel##1{{\escapechar=` \rlap{\hskip.11in\string##1}}}%
	\def\figlabel##1{\noexpand\llap{\string\string\string##1\hskip.66in}}%
	\def\reflabel##1{\noexpand\llap{\string\string\string##1\hskip.37in}}}
\nolabels

%  tagged section numbers

\global\newcount\secno \global\secno=0
\global\newcount\meqno \global\meqno=1
\global\newcount\subsecno \global\subsecno=0

\font\secfont=cmbx12 scaled\magstep1

\def\section#1{\global\advance\secno by1
   \xdef\secsym{\the\secno.}
   \global\subsecno=0
   \global\meqno=1\bigbreak\medskip
   \noindent{\secfont\the\secno. #1}\par\nobreak\smallskip\nobreak\noindent}
%\xdef\secsym{}

\def\subsection#1{\global\advance\subsecno by1
    %\xdef\secsym{\the\subsecno}
\medskip
\noindent
{\bf\the\secno.\the\subsecno\ #1}
\par\medskip\nobreak\noindent}
%\xdef\secsym{}

\def\newsec#1{\global\advance\secno by1
   \xdef\secsym{\the\secno.}
   \global\meqno=1\bigbreak\medskip
   \noindent{\bf\the\secno. #1}\par\nobreak\smallskip\nobreak\noindent}
\xdef\secsym{}

\def\appendix#1#2{\global\meqno=1\xdef\secsym{\hbox{#1.}}\bigbreak\medskip
\noindent{\bf Appendix #1. #2}\par\nobreak\smallskip\nobreak\noindent}

%         equations

\def\eqnn#1{\xdef #1{(\secsym\the\meqno)}%
	\global\advance\meqno by1\eqnlabel#1}
\def\eqna#1{\xdef #1##1{\hbox{$(\secsym\the\meqno##1)$}}%
	\global\advance\meqno by1\eqnlabel{#1$\{\}$}}
\def\eqn#1#2{\xdef #1{(\secsym\the\meqno)}\global\advance\meqno by1%
	$$#2\eqno#1\eqlabel#1$$}

%			 footnotes

\def\myfoot#1#2{{\baselineskip=14.4pt plus 0.3pt\footnote{#1}{#2}}}
%sequentially numbered footnotes
\global\newcount\ftno \global\ftno=1
\def\foot#1{{\baselineskip=14.4pt plus 0.3pt\footnote{$^{\the\ftno}$}{#1}}%
	\global\advance\ftno by1}

%         references

\global\newcount\refno \global\refno=1
\newwrite\rfile

\def\ref{[\the\refno]\nref}
\def\nref#1{\xdef#1{[\the\refno]}\ifnum\refno=1\immediate
	\openout\rfile=refs.tmp\fi\global\advance\refno by1\chardef\wfile=\rfile
	\immediate\write\rfile{\noexpand\item{#1\ }\reflabel{#1}\pctsign}\findarg}
%	horrible hack to sidestep tex \write limitation
\def\findarg#1#{\begingroup\obeylines\newlinechar=`\^^M\passarg}
	{\obeylines\gdef\passarg#1{\writeline\relax #1^^M\hbox{}^^M}%
	\gdef\writeline#1^^M{\expandafter\toks0\expandafter{\striprelax #1}%
	\edef\next{\the\toks0}\ifx\next\null\let\next=\endgroup\else\ifx\next\empty%

\else\immediate\write\wfile{\the\toks0}\fi\let\next=\writeline\fi\next\relax}}
	{\catcode`\%=12\xdef\pctsign{%}}
\def\striprelax#1{}

\def\semi{;\hfil\break}
\def\addref#1{\immediate\write\rfile{\noexpand\item{}#1}} %now unnecessary

\def\listrefs{\vfill\eject\immediate\closeout\rfile
   {{\secfont References}}\bigskip{\frenchspacing%
   \catcode`\@=11\escapechar=` %
   \input refs.tmp\vfill\eject}\nonfrenchspacing}

\def\putrefs{\immediate\closeout\rfile
   {{\bf References}}\bigskip{\frenchspacing%
   \catcode`\@=11\escapechar=` %
   \input refs.tmp }\nonfrenchspacing}

\def\startrefs#1{\immediate\openout\rfile=refs.tmp\refno=#1}

%		and finally, figures:

\global\newcount\figno \global\figno=1
\newwrite\ffile
\def\fig{\the\figno\nfig}
\def\nfig#1{\xdef#1{\the\figno}\ifnum\figno=1\immediate
	\openout\ffile=figs.tmp\fi\global\advance\figno by1\chardef\wfile=\ffile
	\immediate\write\ffile{\medskip\noexpand\item{Fig.\ #1:\ }%
	\figlabel{#1}\pctsign}\findarg}

\def\listfigs{\vfill\eject\immediate\closeout\ffile{\parindent48pt
	\baselineskip16.8pt{{\secfont Figure Captions}}\medskip
	\escapechar=` \input figs.tmp\vfill\eject}}

%%%%%%%%%%%%%%%%%%%%%%%%%%%%%%%%%%%%%%%%%%%%%%%%%%%%%%%%%%%%%%%%%%%%%%%%%%%%%%%
\def\noblackbox{\overfullrule=0pt}
\def\inv{^{\raise.18ex\hbox{${\scriptscriptstyle -}$}\kern-.06em 1}}
\def\dup{^{\vphantom{1}}}
\def\Dsl{\,\raise.18ex\hbox{/}\mkern-16.2mu D} %this one can be subscripted
\def\dsl{\raise.18ex\hbox{/}\kern-.68em\partial}
\def\slash#1{\raise.18ex\hbox{/}\kern-.68em #1}
\def\lspace{}
\def\lbspace{}
\def\boxeqn#1{\vcenter{\vbox{\hrule\hbox{\vrule\kern3.6pt\vbox{\kern3.6pt
	\hbox{${\displaystyle #1}$}\kern3.6pt}\kern3.6pt\vrule}\hrule}}}
\def\mbox#1#2{\vcenter{\hrule \hbox{\vrule height#2.4in
	\kern#1.2in \vrule} \hrule}}  %e.g. \mbox{.1}{.1}
%matters of taste
%\def\tilde{\widetilde}
\def\bar{\overline}
\def\e#1{{\rm e}^{\textstyle#1}}
\def\del{\partial}
\def\curly#1{{\hbox{{$\cal #1$}}}}
\def\curlyD{\hbox{{$\cal D$}}}
\def\curlyL{\hbox{{$\cal L$}}}
\def\vev#1{\langle #1 \rangle}
\def\psibar{\overline\psi}
\def\lform{\hbox{$\sqcup$}\llap{\hbox{$\sqcap$}}}
\def\darr#1{\raise1.8ex\hbox{$\leftrightarrow$}\mkern-19.8mu #1}
\def\half{{\textstyle{1\over2}}} %puts a small half in a displayed eqn
\def\roughly#1{\ \lower1.5ex\hbox{$\sim$}\mkern-22.8mu #1\,}
\def\MSbar{$\bar{{\rm MS}}$}
%%%%%%%%%%%%%%%%%%%%%%%%%%%%%%%%%%%%%%%%%%%%%%%%%%%%%%%%%%%%%%
\hyphenation{di-men-sion di-men-sion-al di-men-sion-al-ly}

\parindent=10pt
\parskip=5pt

\def\Tr{{\rm Tr}}
\def\det{{\rm det}}
\def\jump{\hskip1.0cm}
\def\wzw{Wess--Zumino--Witten}
\def\Az{A_z}
\def\Azb{A_{\bar{z}}}
\def\lr{\lambda_R}
\def\ll{\lambda_L}
\def\lrb{\bar{\lambda}_R}
\def\llb{\bar{\lambda}_L}
\font\top = cmbxti10 scaled \magstep1

\def\d{\partial_z}
\def\db{\partial_{\bar{z}}}
\def\rline{{{\rm I}\!{\rm R}}}
\def\tl{t_L}
\def\tr{t_R}
\def\IR{{\hbox{{\rm I}\kern-.2em\hbox{\rm R}}}}
\def\IB{{\hbox{{\rm I}\kern-.2em\hbox{\rm B}}}}
\def\IN{{\hbox{{\rm I}\kern-.2em\hbox{\rm N}}}}
\def\IC{{\hbox{{\rm I}\kern-.6em\hbox{\bf C}}}}
\def\IZ{{\hbox{{\rm Z}\kern-.4em\hbox{\rm Z}}}}
\noblackbox
\def\WZW{Wess--Zumino--Witten}
\font\Bigtitlerm=cmr10 scaled \magstep4
\font\Footfont=cmr10 scaled \magstep1
\preprint{
\vbox{
\rightline{IASSNS--HEP--94/66}
\vskip2pt\rightline{hep-th/9409061}
\vskip2pt\rightline{July 1994}
}
}
\vskip-1.0cm
\title{\Bigtitlerm Heterotic Cosets\myfoot{$^\dagger$}{\Footfont
A
talk given at the {\sl `Trieste Summer School  on High Energy Physics
and Cosmology'}, 29th July 1994.}} \vskip-1.0cm \author{Clifford V.
Johnson\myfoot{$^*$}{\rm Address after 1st September 1994: Joseph Henry
Laboratories, Jadwin Hall,  Princeton University,
Princeton NJ 08544, U~S~A.}}{}{}
\vskip1.0cm \addressline{\sl
School of Natural Sciences}
\addressline{\sl Institute for Advanced Study}
\addressline{\sl Olden Lane}
\addressline{\sl Princeton NJ 08540}
\addressline{\sl U S A}

\abstract

\bigskip

A description is given of how to construct $(0,2)$ supersymmetric
conformal field theories as coset
models. These models
 may be used as non--trivial backgrounds for Heterotic
String Theory. They are realised as a combination  of an anomalously
gauged  Wess--Zumino--Witten model, right--moving supersymmetric
fermions, and left--moving current algebra fermions. Requiring the
sum of the gauge anomalies from the bosonic and fermionic sectors to cancel
yields the final model.  Applications discussed include exact models of
extremal  four--dimensional
charged black holes and Taub--NUT solutions of string theory.
These coset models may also be used to construct  important families of
$(0,2)$ supersymmetric Heterotic String compactifications. The Kazama--Suzuki
models are the left--right symmetric  special case of these  models.
\vskip-1.5cm
%\draft
\date{(Revised September 1994)}

\section{Introduction and Motivation}
My aim here is to show  how to construct non--trivial conformal field
theories with $(0,2)$ supersymmetry. The motivation is clear: It is well
known that in order to obtain the desired $N=1$ spacetime supersymmetry in
heterotic string theory, the minimum requirement is world
sheet $N=2$
supersymmetry. Well, we have heard in the School about the  highly studied
$(2,2)$
conformal field theories and many  of the fascinating facts
about their moduli spaces  (e.g. Mirror Symmetry). However, these models are
in a sense over--specialised examples
of the generic $(0,2)$ supersymmetric conformal field theories which
heterotic string theory demands. In this sense, the task of studying the
moduli space of heterotic string vacua has only just begun. A search for
many  $(0,2)$ models and understanding of their moduli has to begin in
earnest. This talk will describe the construction of  isolated points in
this moduli space. I first describe the general case and end with some
examples and a brief discussion.

\section{(2,2) cosets: Kazama--Suzuki Models}

Coset models were first invented by Bardakci and Halpern\ref\bardhal{K.
Bardakci and M. B. Halpern, Phys. Rev. {\bf  D3} (1971) 2493\semi M. B.
Halpern, Phys. Rev. {\bf D4} (1971) 2398.} and later generalised by Goddard,
Kent and Olive\ref\gko{P. Goddard and D. Olive, Nucl. Phys. {\bf B257} (1985)
226\semi P
Goddard, A Kent and D Olive, Phys Lett {\bf B152} (1985) 88\semi
P Goddard, A Kent and D Olive, Commun Math Phys {\bf 103} (1986) 105.}\ as
algebraic realisations of new conformal systems, `$G/H$' based upon
affine Lie algebras (a special case of Kac--Moody algebras\ref\km{V. G. Kac,
Funct. Anal. App. {\bf 1} (1967) 328\semi R. V. Moody, Bull. Amer. Math. Soc.
{\bf 73} (1967) 217.}\ref\kmmore{V. G. Kac,  {\sl `Infinite-dimensional Lie
Algebras---An Introduction'}, Birkh\"auser, Basel 1983, 2nd Edition Cambridge
University Press, Cambridge 1985.}) for a group $G$ and a subgroup $H$. The
$N=1$
supersymmetric extension was worked out soon after and is based upon
analogous constructions using affine Lie superalgebras\ref\supergko{V. G. Kac
and T. Todorov, Commun. Math. Phys. {\bf 103} (1986) 105.}.
(For a review see ref.\ref\goddardolive{P. Goddard and D. Olive, Int. Jour.
Mod. Phys.
{\bf A1} (1986) 303.}.)

When the space $G/H$ is K\"ahler, it was shown by Kazama and
Suzuki\ref\ks{Y Kazama and H Suzuki, Nucl Phys {\bf B321} (1989) 232\semi
Y Kazama and H Suzuki, Phys Lett {\bf B216} (1989) 112.}\
using an algebraic construction that  $N=1$  is promoted
to an $N=2$ supersymmetry. This realises a large family of $(2,2)$
models, of which the $N=2$ minimal models (used for example by Gepner in
his construction of non--trivial heterotic string vacua) are the simplest
case (they are realised as $SU(2)/U(1)$). The most straightforward examples
are the `Hermitian Symmetric
Spaces'.

For  many reasons it is advantageous to have a Lagrangian definition of a
conformal field theory which realises the algebraic structures as a field
theory. It is very often a powerful supplement to the algebraic description.
The {\sl Gauged Wess--Zumino--Witten Model} is the appropriate device to
use.

\section{(2,2) Cosets as Gauged Wess--Zumino--Witten Models}

An action for a conformal field theory with all of the
algebraic structures of the Kazama--Suzuki models is:
\def\DG{{g^{-1}\d g}}\def\DBG{{\db gg^{-1}}}
\eqn\SGWZW{\eqalign{I^{(2,2)}& =I_{WZW}(g)+I(g,A)+I_F(\Psi_L,\Psi_R,A)=\cr
&-{k\over4\pi}\int_{\Sigma} d^2z\,\, \Tr[g^{-1}\d g\cdot g^{-1}\db
g]\cr &-{i\over12\pi}\int_B d^3\sigma\,\,
\epsilon^{ijk}\Tr[g^{-1}\partial_ig\cdot g^{-1}\partial_jg\cdot g^{-1}
\partial_kg]\cr
&+{k\over2\pi}\int_\Sigma d^2z\,\,\Tr[\Az\DG-\Azb\DBG+\Azb g^{-1}\Az
g-\Az\Azb]\cr
&+{i\over4\pi}\int_\Sigma d^2z\,\,\Tr[\Psi_+{\cal D}_{\bar
z}\Psi_++\Psi_-{\cal D}_z \Psi_-]
}}
where $\Sigma=\partial B$ and
$$
\eqalign{g\in& G;\,A^a\in{\rm Lie}H;\cr
\Psi_\pm\in&{\rm Lie}G-{\rm Lie}H,\,\,\,{\cal D}_a\equiv \partial_a+
[A_a,\,\,\,]}
$$
and we have gauged the group invariance $$\eqalign{g\to&hgh^{-1},\cr
{\rm where}\,\,\,h\in& H.}$$
This action has an $N=1$ supersymmetry:
\eqn\Neqone{\eqalign{
&\delta g=i\epsilon_-g\Psi_++i\epsilon_+\Psi_-g\cr
&\delta \Psi_+=\epsilon_-(1-\Pi_0)\cdot(g^{-1}{\cal D}_zg-i\Psi_+\Psi_+)\cr
&\delta \Psi_-=\epsilon_+(1-\Pi_0)\cdot({\cal D}_{\bar
z}gg^{-1}+i\Psi_-\Psi_-)\cr
}}
where $\Pi_0$ is the orthogonal projection of Lie$G$ onto Lie$H$.

Now, just as in the algebraic construction of Kazama and Suzuki, an $N=2$
supersymmetry arises from this $N=1$ when the space $G/H$ is K\"ahler. I
will not dwell on this further here, save to note that this action was first
studied in this context  by Witten\ref\ed{E Witten, Nucl Phys {\bf B371}
(1992) 191.}\
and Nakatsu\ref\Nakatsu{S Nakatsu, Prog Theor Phys {\bf 87} (1992) 795.}.
Witten used this action (after twisting) to
do explicit calculations in certain topological field theories.  The
explicit $N=2$ transformations are written down
in ref.\ref\mans{M Henningson, Nucl Phys {\bf B413} (1994) 73\semi
M Henningson, Institute for Advanced Study preprint IASSNS--HEP--94/13,
hep-th/9402122.}\ for example and there Henningson uses the  models to study
 important properties of the Kazama--Suzuki models which are more
easily accessible via field theoretic methods. This includes  a
demonstration of mirror symmetry for the Kazama--Suzuki models and a
calculation of the elliptic genus for the $N=2$ minimal models.

Note by the way that the bosonic and fermionic sectors in \SGWZW\ are
consistent models. In particular, the bosonic sector of the gauged \WZW\
model is of course a consistent model realising the  bosonic
cosets\ref\gauging{See for example D Karabali, Q--H Park, H Schnitzer and
Z Yang, Phys Lett {\bf B216} (1989) 307.}\
and the action for the  chiral fermions, when written in this `coset'
basis, is just a simple minimal coupling to the gauge fields\ref\rohm{R
Rohm, Phys Rev {\bf D32} (1985) 2849.}.
The chiral anomalies which potentially arise from this coupling exactly
cancel due to the identical nature of the left and right fermion couplings.
The anomalies contribute with opposite sign but equal magnitude.

\section{(0,2) Cosets: Potential Problems and a Solution}

\noindent
(1) To get a $(0,2)$ conformal field theory, we need to remove the left
$N=2$.
Simply deleting or changing the couplings of the left moving fermions to
the gauge fields would certainly do this for us, without spoiling the
right--moving $N=2$. The only problem is that this procedure is bound to
produce anomalies. The right--movers' chiral anomaly will either have
nothing to cancel against (if we deleted the left--movers), or will not
completely cancel (if we changed the couplings of the left--movers to
spoil the third symmetry in \Neqone).

\noindent
(2)
For many other reasons (as will be illustrated later), it would also be nice
to gauge other symmetries of the WZW model. To get a consistent model,
one has to gauge a restricted class of subgroups of the full $G_L\times
G_R$ symmetry which exists for the basic WZW. These are called
`anomaly--free' subgroups, mainly because one of the first uses of this
type of model (in higher dimensional gauge theories) was to study the
structure of anomalies\ref\anomalyref{See for example the book {\sl
`Current Algebra and Anomalies'}, S B Treiman, E Witten, R Jackiw and B
Zumino, World Scientific, Singapore 1986.}\ by deliberately
studying anomalous subgroups, and then letting the Wess--Zumino term
produce {\sl classically} the familiar quantum gauge anomalies.
Since Witten's paper on the use of the Wess--Zumino term to define a
conformally invariant sigma model in two  dimensions---the \WZW\
model---most of
the efforts involving them in 2D, including their gauged versions, have
made sure that there are no anomalies. This is simply because the model
would not correctly reproduce the coset algebra---it would not  be
conformally invariant, in general.

Given the language I just used to describe the problems we would like to
solve, it is clear that a solution presents itself in the form of {\sl
cancelling the anomalies against one another.} If we arrange things
correctly, this will work. In the next section I describe just how to do
this.

\section{Anomalies}

There are anomalies arising from three sectors now. The classical anomaly
from the WZW  and the chiral anomalies at one--loop from each chirality
of fermion. I will discuss each in turn.

\bigskip

\noindent
{\bf The WZW anomalies.}

\bigskip

In general gauging the following  symmetry of the WZW model
$$\eqalign{&g\to h_1gh_2^{-1}\cr
{\rm for}\,\,\,&(h_1,h_2)\in(H_L,H_R)\subset(G_L,G_R)}$$ is anomalous. This
simply means that one cannot write down an extension of the
WZW model which promotes this symmetry to a local invariance: There will
always be terms which spoil gauge invariance. (This is because of the
Wess--Zumino term; the metric term may be simply minimally coupled.)

Knowing that we will get an anomaly, let us choose to write {\sl some}
gauge extension such that under gauge transformations the `anomalous'
piece does not depend upon the group element
$g$. This results in the anomalous piece taking
the form of the standard 2D chiral anomaly. The  {\sl unique} action
is\ref\edagain{E Witten, Commun Math Phys {\bf 144} (1992) 191.}:
\eqn\extend{\eqalign{I^{G_k}_{GWZW}(g,A)&=-{k\over4\pi}\int_\Sigma\,\,
d^2z \,\,\Tr[g^{-1}{\cal D}_zg\cdot g^{-1}{\cal D}_{\bar z}g]\cr
&-{i\over12\pi}\int_B d^3\sigma\,\,
\epsilon^{ijk}\Tr[g^{-1}\partial_ig\cdot g^{-1}\partial_jg\cdot g^{-1}
\partial_kg]\cr
&-{k\over4\pi}\int_\Sigma A^a\wedge\Tr[t_{a,L}\cdot
dgg^{-1}+t_{a,R}g^{-1}dg]\cr
&-{k\over8\pi}\int_\Sigma A^a\wedge A^b\Tr[t_{a,R}g^{-1}t_{b,L}g-
t_{b,R}g^{-1}t_{a,L}g].}}
Under the infinitesimal variation
$$\eqalign{g\to&g+ \sum_a \epsilon_a (t_{a,L}g-gt_{a,R})\cr
\Az^a\to& \Az^a+{\cal D}_{z}\epsilon^a\cr
\Azb^a\to& \Azb^a+{\cal D}_{\bar z}\epsilon^a,\cr
}$$
the variation is
\eqn\vary{
\eqalign{
\delta I(g,A)&={k\over4\pi}\Tr[t_{a,R}\cdot t_{b,R}-t_{a,L}\cdot
t_{b,L}]\int_\Sigma d^2z \epsilon^{(a)}F^{(b)}_{z\bar z}\cr
{\rm where}\,\,\,&t_{a,L(R)}\in {\rm Lie}H_{L(R)}.
}}
Notice in particular that for the popular diagonal gaugings of WZW models
this variation is zero and the action reduces to the familiar one.

\bigskip

\noindent
{\bf The right movers}

\bigskip

As mentioned before, it is sufficient to minimally couple the coset fermions
to the gauge fields:
\eqn\IFR{\eqalign{I_F^R(\Psi_R,A)=&{k\over4\pi}\int_\Sigma\,\,
i\Tr[\Psi_R{\cal D}_{\bar z}\Psi_R]\cr
{\rm where}\,\,{\cal D}_{\bar z}\Psi_R&=\db\Psi_R+\sum_a
\Azb^a[t_{a,R},\Psi_R],\,\,\Psi_R\in{\rm Lie}G-{\rm Lie}H.}}

There are $D=$dim$G-$dim$H$ fermions $\psi^i_R$ in $\Psi_R$, all coupled
with charges derived from the generators $t_{a,R}$.
The chiral anomalies appear at one loop and  are:
\eqn\Ranomalies{{D\over4\pi}\Tr[t_{a,R}\cdot t_{b,R}]\int_\Sigma d^2z
\epsilon^{(a)}F^{(b)}_{z\bar z}.}
(Note here  the absence of $k$, which plays the role of $1/\hbar$. This
 really is a one loop effect.)

\bigskip

\noindent
{\bf The left movers}

\bigskip

Let us couple into the model some left movers. Let us add
$D=$dim$G-$dim$H$ of
them (a good choice, as we will see later) with arbitrary couplings. To
be precise, arrange them into a fundamental vector
$\Lambda_L=\{\lambda^i_L\}$ of the
group \def\SOD{SO({\rm dim }G-{\rm dim}H)}
$SO(D)_L$ which acts on them as a global symmetry, and minimally couple
them to the
$H_L$ subgroup with generators $Q_{a,L}$ in this fundamental representation:
\eqn\IFL{I_F^L(\lambda^i_L,A)={k\over4\pi}\int_\Sigma\,\,
i\Lambda^T_L(\d+\sum_a\Az^aQ_{a,L})\Lambda_L.} (Here $\tilde\Tr$ is the
trace in the fundamental representation of $SO(D)$.)

Their chiral anomalies
appear at one loop and are:
\eqn\Lanomalies{-{1\over4\pi}{\tilde\Tr}[Q_{a,L}\cdot Q_{b,L}]\int_\Sigma
d^2z \epsilon^{(a)}F^{(b)}_{z\bar z}.}
(Note again the absence of $k$. Also note the minus sign relative to
\Ranomalies, due to the oppisitte chirality.)

So if we add together the three actions \extend,\IFR\ and \IFL, we get
a gauge invariant  model if we ensure
that all of the  anomalies (classical and
quantum) cancel: \eqn\cancelone{k\Tr[t_{a,R}\cdot
t_{b,R}-t_{a,L}\cdot t_{b,L}]+\Tr[t_{a,R}\cdot
t_{b,R}]-{\tilde\Tr}[Q_{a}\cdot Q_{b}] =0.}

Our model has $(0,2)$ supersymmetry as advertised (because we have not
touched the right--moving sector), and is conformally invariant.

Well, our model is gauge invariant when we take into account the
one--loop effects, but we still have not written a {\sl classically}
gauge invariant action. This means that we cannot truly carry out
procedures like path--integral quantisation, etc. We have not quite
achieved our goal of  a Lagrangian realisation of a $(0,2)$ conformal
field theory.

The answer is to {\sl bosonize} the fermions. The bosonic action
equivalent to $I_R^F+I_L^F$ is {\sl classically} anomalous. It is a
theory of $D/2$ real bosons with the same anomalies
as above.

\section{Bosonisation}

I will give specific examples later, where I have worked out the
bosonisation by hand in some abelian cases. After a little thought,
however, it is clear  once one realises that  a
classically anomalous bosonic theory equivalent to an anomalous
fermionic theory is to be found, it might be that  the bosonic theory is
something like another anomalously gauged WZW.

Note that before gauging there are $D$ free fermions on the left and right.
They therefore carry a global $SO(D)_L\times SO(D)_R$ symmetry. Witten
showed in ref.\ref\ed{E Witten, Commun Math Phys {\bf 92} (1984) 455}\ that
this system of free fermions
is equivalent to a \WZW\ model based on $SO(D)$ at level 1. Considering
what we saw about WZW anomalies in earlier section it is clear that the
classically anomalous bosonic theory equivalent to the fermionic theory
is just this $SO(D)$ WZW gauged anomalously with different embeddings of
$H$ in $SO(D)$ on the left and on the right:
\def\tg{{\tilde g}}
\def\th{{\tilde h}}

$$
\eqalign{{\tilde g}\to& {\tilde h}_1{\tilde g}{\tilde h}_2\cr
{\rm for}\,\,\,{\tilde g}\in& SO(D)\,\,\,{\rm and}\cr
(\th_1,\th_2)\in&(H_L,H_R)\subset(SO(D)_L,SO(D)_R)
}
$$
Let the $(H_L,H_R)$ be generated by $(Q_{a,L},Q_{a,R})$. Choose the
$Q_{a,R}$ such that when acting on the $\psi_R^i$'s in the fundamental
representation of $SO(D)$ they are equivalent to the $t_{a,R}$ acting on
the $\psi^i_R$ in the coset fermion $\Psi_R\in{\rm Lie}G-{\rm Lie}H$.
This will ensure that the right moving fermions are correctly coupled and
preserve the (now hidden) $N=2$ on the right.

Then the bosonic action equivalent to the interacting fermions is just an
action of the form \extend\ (with level 1), which yields the classical
anomalies:
$$
{1\over 4\pi}{\tilde\Tr}
[Q_{a,R}\cdot Q_{b,R}-Q_{a,L}\cdot Q_{b,L}]
\int_\Sigma d^2z \,\,\epsilon^{(b)}F_{z\bar z}^{(a)}.
$$
So cancelling this against the anomaly of the $G/H$ bosonic model
(and recalling from the above paragraph that
${\tilde\Tr}[Q_{a,R}\cdot Q_{b,R}]=D\Tr[t_{a,R}\cdot t_{b,R}]$), we
recover \cancelone\ as the condition for a consistent model.

\section{(0,2) Cosets as Gauged \WZW\ Models}

So finally we can write a classically gauge invariant analogue of
\SGWZW\ which realises a $(0,2)$ conformal field theory as a gauge
invariant action written as the sum of two gauged \WZW\ models which are
separately anomalous:
\eqn\final{I^{(0,2)}=I_{GWZW}^{G_k}(g,A)+I_{GWZW}^{SO(D)_1}(\tg,A),}
where $ D={\rm dim}G-{\rm dim}H$.

The heterotic coset is realised as:
$\left[G_k\times SO(D)_1\right]/ H$ with the gauged symmetry:
$$
\eqalign{g\to&h_2gh_1^{-1}\cr
\tg\to&\th_2\tg\th_1^{-1}\cr
\hbox{\rm subject to}\,\,\,k\Tr[t_{a,R}\cdot t_{b,R}-t_{a,L}\cdot t_{b,L}]+&
D\Tr[t_{a,R}\cdot t_{b,R}]-{\tilde \Tr}[Q_{a,L}\cdot Q_{b,L}]=0.}
$$
Note that $h_1$ and $\th_1$ are chosen so as to recover right--moving
supersymmetry  in the fermion picture.

Note that in \final\ the gauge extensions to each WZW (written using
\extend) are generally not  gauge
invariant, but together they are because of the anomaly equation above.
In the special case of $h_2=h_1$ and $\th_2=\th_1$, they are each
separately gauge invariant extensions, the anomaly equation is trivially
satisfied, and we recover the $(2,2)$ case.
{\sl In this sense, the $(2,2) $ models can now be seen as a special case
of a more general class of $(0,2)$ models.}

\section{Some examples.}
I originally used these ideas to study some particular
cases\ref\paperone{C V Johnson, Institute for Advanced Study preprint
IASSNS--HEP--94/20, to appear in Phys. Rev. {\bf D}. hep-th/9403192}.
The prototype model for this construction is the `monopole theory' of
Giddings, Polchinski and Strominger\ref\gps{S Giddings, J Polchinski and A
Strominger, Phys Rev {\bf D48} (1993) 5784. hep-th/9405083} (GPS). It
is a conformal field theory of a heterotic string in a Dirac monopole
background of charge $Q$ on a two--sphere of radius of order $Q$.
GPS described it as an asymmetric orbifold
of $SU(2)$. Here, described as a heterotic coset, it is based upon an $SU(2)$
 WZW with
the $U(1)$ subgroup of the right $SU(2)$ gauged. Adding supersymmetric
right  movers and left movers of charge $Q$ gives an anomaly equation
$k=2(Q^2-1)$. Bosonising the fermions it is possible to correctly
determine the quadratic terms in the gauge fields which
turns out to only depend upon $Q$. After integrating
out the gauge fields, and correctly re--fermionising the action, the
heterotic sigma model describing the above system is recovered.
This is described in detail in ref.\paperone.
As pointed out by GPS, the tensor product of this model with a supersymmetric
 $SL(2,\rline)/U(1)$
2D black hole coset\ref\bh{E Witten, Phys Rev {\bf D44} (1991) 314.}\ yields
a 4D solution which is the extremal
limit of the magnetically charged dilaton black hole of Gibbons, Maeda
 and Garfinkle, Horowitz and Strominger\ref\QblackHole{G W Gibbons and K
Maeda, Nucl Phys {\bf B298} (1988) 741\semi
D Garfinkle, G T Horowitz and  A Strominger, Phys Rev {\bf D43} (1991)
3140, erratum Phys Rev {\bf D45} (1992) 3888.}.

Notice that in the construction I described  for the monopole
theory, one cannot have a charge $Q=0$ solution, as then the anomaly
equation would not be satisfied. After a little thought, it is clear that
there is a quick way out of this problem: simply gauge $g\to hg$ instead
and  keep everything else the same. Then the sign of the WZW anomaly
changes and the condition $k=2(1-Q^2)$ should now be satisfied. Now it is
possible to get a $Q=0$ solution. (In constructing their neutral
solution in their paper, Giddings, Polchinski and Strominger arrive
at this simple modification in an equivalent way. This is indeed the same
solution).
Now naively, the interpretation of the model  would be as a heterotic string on
a neutral two--sphere background. However, it is easy to see that this is
wrong. The problem of incorrectly identifying
the two--sphere as the background manifold for small $Q$ has its roots in the
fact that the final form of the metric for the model  is obtained by
integrating out the constraining 2D gauge fields, a process which is well
defined only for large $Q$. which is equivalent to small $\alpha^\prime$, or
large $k$. Here, the neutral solution has  $k=2$, and no sensible metric
interpretation may be made of the target space via perturbation theory, as all
length scales (in units of $\alpha^\prime$) contribute equally to the
$\beta$--function equations.

The most obvious application of this construction at the time was to find
more general 4D solutions which were dyons (both magnetic and electric
charge). Applying this construction to general gaugings of $SL(2,\rline)$
was carried  out in ref.\paperone, yielding at leading order the known
2D charged black hole heterotic string solutions of McGuigan,
Nappi and Yost\ref\charged{M D McGuigan, C R Nappi and S A Yost, Nucl
Phys {\bf B375} (1992) 421. hep-th/9111038.}, and 4D dyonic solutions were
defined by tensor product with the
GPS theory. At about the same time, Lowe and Strominger wrote a
paper\ref\lowe{D A Lowe and A Strominger,  hep-th/9403186.}\
about 4D dyons which were defined by tensoring the GPS theory with an
asymmetric orbifold of $SL(2,\rline)$. This asymmetric orbifold may be
realised as one of the $SL(2,\rline)$ heterotic cosets described in
ref.\paperone.

Instead of tensor products of these 2D theories,
it is possible to obtain 4D dyon solutions which
are not tensor products by gauging
 (for example) a $U(1)\times U(1)$ subgroup of $SL(2,\rline)\times SU(2)$
embedded non--trivially such that the action of the $U(1)$'s was shared
among the two parent groups. In this way I obtained in ref.\paperone\ a 4D
dyon with a throat
with a non--trivial mixing of the angular and radial coordinates\foot{From
the form of the low energy solution, I conjectured that this was
a dyonic, axionic analogue of the Taub--NUT solution of General Relativity.
With Myers\ref\papertwo{C V Johnson and R C Myers, Institute
For Advanced Study Preprint IASSNS--HEP--94/50 and McGill Preprint
McGill/94--28, to appear in Phys. Rev. {\bf D}. hep-th/9406069.}\ this
conjecture was later confirmed  by explicitly
generating
the full solution by using first $T$ and then $S$ duality transformations
on the GR solution and then extremising it. Kallosh, Kastor Ortin and Torma
\ref\kallosh{R Kallosh, D
Kastor, T Ortin and T Torma, hep-th/9406059.} also constructed this
solution at around the same time.}. It would have been difficult to construct
such a non--trivial solution
as an exact conformal field theory
without the use of the heterotic coset
technique.

\section{Future directions}

There are a {\sl huge} number of avenues opened by allowing such freedom to
gauge any subgroup of the WZW model's symmetries, obtaining consistency by
adding  heterotic fermions. I cannot list all of the things
which occur to me here, but the general point is that it allows one to
consider leaving  important WZW symmetries untouched, which in turn
leaves certain spacetime symmetries intact. In the simple GPS monopole
example, or in the even simpler example of an uncharged 2--sphere
in the last section, leaving the $SU(2)_L$ (or $SU(2)_R$) action untouched
meant that a simple spacetime spherically symmetric system was obtained
from an $SU(2)$ WZW.

This type of freedom will certainly lead to many more interesting
heterotic string backgrounds. The search for more 4D cosmological
heterotic  string
backgrounds seems a promising area to apply this technique to.

Of great interest is the problem of calculating the spectrum and partition
function for these models. This will be of course a highly non--trivial
combination of right $N=2$ characters and general $N=0$ characters. It is
a hard problem to discover the heterotic modular invariant combinations
algebraically of course (see e.g. ref.\ref\gannon{See for example T
Gannon, Nucl Phys {\bf B402} (1993) 729. hep-th/9209042. }), but there are
promising signs that their consistent
description as a WZW as described in this talk may provide some guidance.
Work is in progress on this and related matters with Berglund, Kachru and
Zaugg\ref\paperthree{P Berglund, C V Johnson, S Kachru and P Zaugg, in
preparation.}.

The problem of starting to map out the  moduli space of $(0,2)$ models
can be
attacked successfully by studying the marginal perturbations of these
models. This is of course much easier when there exists a Lagrangian
description of the type constructed here. Such marginal perturbations
would help to find the geometrical interpretation of the nieghbourhoods of
these models, in the case of their use as string compactifications.

Marginal perturbations
would also represent interesting geometrical freedom in some 4D
solutions, where they correspond to such
 processes as widening the throat of some of the extremal solutions of
the type mentioned in the last section, connecting onto the
asymptotically flat 4D exterior solution\gps.

There are of course many more questions which need to be
answered about the moduli
space of  $(0,2)$ conformal field theories.
I hope that this construction may go some way to help to answer them.

\vfill\eject

{\bf Acknowledgements}
\bigskip

I would like to thank all of the Staff, Organisers  and Participants at
this School and Workshop for making it such a fruitful and enjoyable
experience for me.
I appreciate the hospitality of the International
Centre For Theoretical Physics during my stay here, and where many of the
details of this talk were worked out.

I would also like once again to
thank Ed Witten for originally pointing out to me over a year ago
that the GPS monopole theory might be described by a construction
of the type described in this talk.

This work is supported by an EPSRC (UK) NATO Fellowship.

\listrefs
 \bye